\documentclass[%
reprint,
superscriptaddress,
amsmath,amssymb,
aps,
prl,
]{revtex4-2}

\usepackage{graphicx}
\usepackage{array}
\usepackage{dcolumn}
\usepackage{bm}
\usepackage{listings}
\usepackage{scalefnt}
\usepackage{xcolor}
\usepackage{booktabs}
\usepackage{multirow}
\usepackage{tensor}
\usepackage{hyperref}
\usepackage{mathtools}
\hypersetup{
    colorlinks,
    linkcolor={blue!80!black},
    citecolor={blue!80!black},
    urlcolor={blue!80!black}
}
\usepackage{color}
\usepackage{physics}
\usepackage[pagewise]{lineno}
\usepackage{import}

\usepackage{lipsum}

\def\noteC#1{\textbf{\color{green}}} 

\newcommand{\diag}{\mathrm{diag}}

\begin{document}

\title{Coarse-grained quantum state tomography with optimal POVM construction}

\author{Donghun Jung}
\affiliation{Center for Quantum Information, Korea Institute of Science and Technology, Seoul, 02792, Republic of Korea}
\affiliation{Department of Physics, Sungkyunkwan University, Suwon 16419, Republic of Korea}
\author{Young-Wook Cho}
\affiliation{A*STAR Quantum Innovation Centre (Q.Inc), Institute of Materials Research and Engineering (IMRE), Agency for Science, Technology and Research (A*STAR), 138634 Singapore}
\author{Yosep Kim}
\email{yosep9201 gmail.com}
\affiliation{Department of Physics, Korea University, Seoul, 02841, Republic of Korea}
\author{Junghyun Lee}
\email{jh\_lee@kist.re.kr}
\affiliation{Center for Quantum Information, Korea Institute of Science and Technology, Seoul, 02792, Republic of Korea}

\date{\today}
\begin{abstract}
    Constructing an integrated large-scale qubit system of realistic size requires addressing the challenge of physical crowding among qubits. This constraint poses an issue of coarse-grained (CG) measurement, wherein information from the multi-qubit system is collectively gathered. In this work, we introduce a novel approach to reconstruct the target density matrix from a comprehensive set of Positive Operator-Valued Measures (POVM) using a Parameterized Quantum Circuit (PQC) under the constraint of CG measurement. 
    We improve the robustness and stability of CG quantum state tomography (QST) by optimizing the POVM set to achieve a generalized symmetric informationally complete (GSIC) POVM through maximization of the von Neumann entropy. 
    This optimized construction of CG-POVMs is scalable to an $N$-qubit system. We further discuss a more efficient construction of $N$-qubit CG-QST without exponential increases in two-qubit gates or circuit depth per measurement.
    Our scheme offers a viable pathway towards a detector-efficient large-scale solid-state embedded qubit platform by reconstructing crucial quantum information from collective measurements.
\end{abstract}
\maketitle





\subsection{Introduction}
Quantum technology relies on precise measurement and characterization of quantum systems,
employing methodologies like quantum state tomography (QST)\cite{nielsen2010quantum, RevModPhys.81.299}, quantum process tomography (QPT)\cite{PhysRevLett.93.080502, PhysRevA.87.062119, PhysRevA.77.032322, kim2018direct, kim2020universal}, and randomized benchmarking\cite{dankert2009exact, emerson2005scalable}.
Despite its perceived inefficiency in large-scale qubit systems,
QST still plays a crucial role in validating quantum devices, evaluating the performance of specific quantum gates, and pinpointing errors.
Its widespread adoption reflects its practical utility as a common technique
for inferring the state of a quantum system through strategic measurements and analysis\cite{altepeter20044, james2001measurement, hou2016full, teo2011quantum, qi2017adaptive}.
\begin{figure}[h!]
    \includegraphics[width=0.38\textwidth]{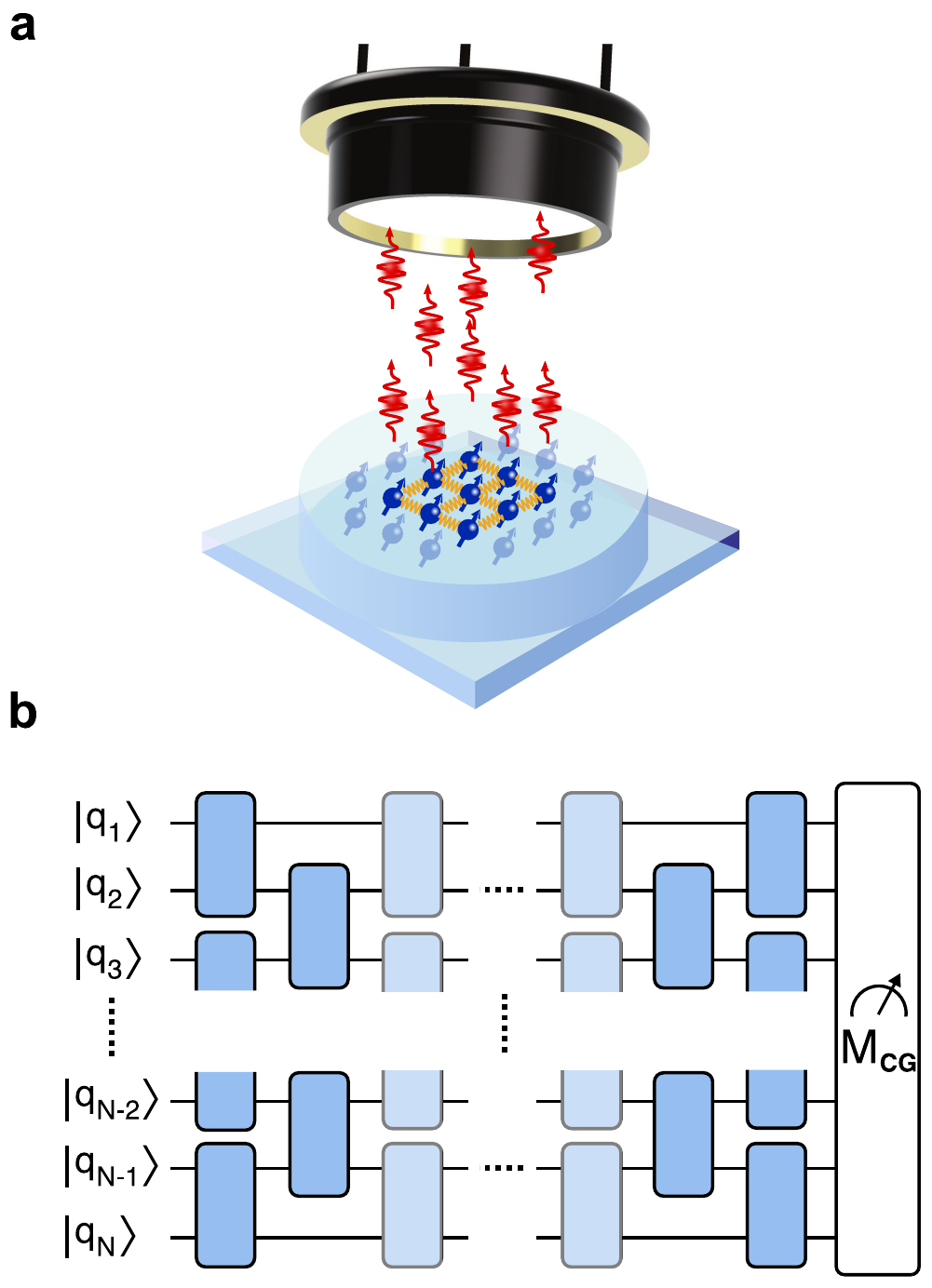}
    \caption{
        (a) Schematic illustration of a collective CG measurement for a multi-qubit system. A single detector collectively captures the total qubit state information, distinguishing only binary outcomes, indicative of the presence or absence of a signal. The qubits are closely located, so that two-qubit interaction is present.
        (b) A brick-wall-shaped parameterized quantum circuit used to establish CG-POVM operators for an $N$-qubit system.
        The circuit comprises layers of blue-colored gate blocks, consisting of parametrized single and two-qubit gates.
    }
    \label{fig:fig1}
\end{figure}

The QST necessitates the preparation of an ensemble of identical quantum systems, each subjected to an informationally complete set of positive operator valued measures (IC-POVMs).
However, the probabilistic outcomes of individual POVM measurement inherently yield limited information, often overshadowed by statistical errors originating from the measuring devices\cite{RevModPhys.68.801, RevModPhys.82.1155}. Therefore, there has been a great interest to find the most noise-resilient and efficient set of measurements for QST among various choice of IC-POVM sets under specifically given conditions. In the class of projective measurements, mutually unbiased bases (MUB)\cite{klappenecker2004constructions, brierley2009all} provide the most efficient information extraction per measurement, facilitating efficient and accurate QST for certain classes of states. However, MUBs for QST are often overcomplete. On the other hand, in the class of generalized POVM measurements, symmetric, informationally complete (SIC) POVMs\cite{PhysRevLett.126.100401, rastegin2014notes} are known as optimal QST measurements with minimal set of POVMs, providing efficient and noise resilient QST especially under noisy environments.

While SIC-POVM based QST is known to be optimal in general cases, the existence of SIC POVMs for all dimensions is still an open question. 
Furthermore, in practice, stringent conditions for SIC-POVMs pose difficulties to implement SIC-POVM based QST. 
For instance, every SIC-POVM element must be rank one, which may not be straightforward to be implemented with noisy measurement apparatus. 
Recently, a generalized concept of SIC POVM has been introduced by relaxing the rank condition. 
Notably, the general SIC-POVMs (GSIC-POVMs)\cite{yoshida2022construction,PhysRevA.90.032309} are shown to exist in all dimensions.
And various properties for GSIC-POVMs have been studied, including uncertainty relations\cite{Huang_2023, rastegin2013uncertainty, PhysRevA.103.042205}, entanglement detections\cite{xi2016entanglement, PhysRevE.107.054134, PhysRevA.91.012326}, and the quantum state tomography\cite{PhysRevA.90.032309}. 
Especially, GSIC POVMs are known to be optimal for QST under the given arbitrary rank of POVM elements.

Although optimal QST strategies for different measurement constraints have been extensively studied, a common assumption is that the physical number of detectors is at least great or equal to the number of qubits to be characterized assuming the ability of individual measurements on each qubit. A totally different scenario should be considered in a highly integrated large-scale qubit system, where the individual state detection is not available anymore. In this paper, we address the problem by introducing a coarse-grained (CG) measurement. In our scenario, we assume that only a single physical detector and adjacent qubit interactions are available as shown in Figure \ref{fig:fig1}a. Under this stringent but realistic CG measurement constraint, we propose a strategic method to characterize a multi-qubit quantum system by optimizing a parameterized quantum circuit. Interestingly, we find that our strategy naturally leads us to the implementation of GSIC POVMs, providing noise-resilient and efficient QST. Note that approaches to find GSIC POVMs were theoretically introduced, but the implementation of GSIC POVMs under specific conditions was not considered yet.

The main part of this letter is organized as follows:
First, we propose a scheme for constructing CG-POVM operators with maximized QST fidelity,
aiming to approach a GSIC POVM set\cite{yoshida2022construction,  PhysRevA.90.032309}.
We adopt von Neumann entropy as a metric to maximize the unbiasedness of the set of CG-POVM bases,
thereby enhancing resilience against statistical noise during the density matrix reconstruction\cite{teo2011quantum, yoshida2022construction, PhysRevA.90.032309}.
Next, we present a circuit resource-efficient method for constructing CG-POVM operators, utilizing a two-qubit system as a toy model, while relaxing the GSIC condition.
The strategy to minimize the use of two-qubit gates and the circuit depth in the context of CG measurement-based QST is discussed.
This is pertinent for the practical implementation of our scheme,
as the fidelity of CG-POVM operators is particularly sensitive to the number of two-qubit gates as well as the circuit depth employed.
Finally, to access the scalability, we extend our optimal CG-POVM construction methodology to an $N$-qubit system with reduced circuit layer and depth.
By employing a brick-wall-shaped parameterized quantum circuit ansatz with gate blocks (Fig.\ref{fig:fig1}b),
we demonstrate the feasibility of $N$-qubit CG-QST without encountering an exponential rise in the number of two-qubit gates or circuit depth per CG-POVM.
Here, we utilize the parameterized Ising coupling gate as a two-qubit gate, where only nearest-neighbor qubits can be coupled\cite{PhysRevA.67.012317, PhysRevA.109.032605}.

\subsection{Coarse-Grained Measurement Background}
In the general QST protocol designed for a multi-qubit system, the assumption is based on the individual readout of each qubit state, projected onto a set of POVM bases formulated from the POVM operators. This involves varying the measurement projection basis of each qubit, often using the Pauli basis, and measure their correlations. The expectation values of these POVM bases are then extracted to reconstruct the system's complete density matrix. However, in the context of coarse-grained (CG) measurements, a single detector captures the aggregate information of the entire system, distinguishing only between two states, such as the presence or absence of signals.(Fig.\ref{fig:fig1}a)\cite{}.
CG measurement becomes prevalent in scenarios involving multi-qubits embedded within a solid-state system,
where the proximity between qubits is essential for constructing efficient two-qubit interactions\cite{dolde2013room, lee2023dressed}.
Individual readout of each qubit presents significant challenges,
attributed to constraints like the optical diffraction limit\cite{dolde2013room} or the collective nature of electrical conductance measurements\cite{dots5coherent}.
Even with super-resolution imaging techniques\cite{pfender2014single, arai2015fourier, Jaskula:17}, scalability remains limited due to system complexities.


In CG measurement, information about the entire multi-qubit system is collectively detected
and the detailed information about their correlations for each qubit's distinct bases is now embedded
in the total event count of correlation-independent projection bases.
The mathematical representation of the CG measurement operator $M_{CG}$ for the entire $N$-qubit system is as follows:
\begin{equation}
    M_{CG} = \frac{1}{N}\left( \sum_{i=1}^{N} I_{2}^{\otimes i-1} \otimes \ket{0}\bra{0} \otimes I_{2}^{\otimes N-i} \right)
\end{equation}
Here, $I_{2}$ denotes the $2 \times 2$ identity operator.
The CG measurement process can be represented by a CG-POVM comprising two positive semi-definite observable operators $\left\{ M_{CG}, I_{2^{N}} - M_{CG} \right\}$.
Due to the dilution of correlative information under CG measurement,
it becomes impossible to contruct an informationally complete set of CG-POVM operators generated solely by using single-qubit gates.
Meanwhile, additional construction of CG-POVM operators $\Omega^{k} = G^{k\dagger} M_{CG} G^{k}$, forming the set of CG-POVM bases $\{ \Omega^{k}, I_{2^{N}} - \Omega^{k} \}$,
can be achieved by implementing a unitary gates $G^{k}$, involving two-qubit gates.
This approach effectively facilitates the reconstruction of correlated measurements,
even under the condition of incoherent collective summation of multi-qubit state information.
For a specific example of CG-POVM for two-qubit case for QST, please see Supplementary Information.





\subsection{Construction of CG-POVM towards GSIC}
We approach the problem of CG-QST as solving a linear equation problem.
Consider a simplified model of a two-qubit system. The density matrix $\rho$ can be projected onto a 16-dimensional vector $\Theta$
by selecting a complete set of orthonormal basis in the corresponding Hilbert-Liouville space $\{ \Gamma_{i} \}_{i=1}^{16}$, satisfying $\Tr ( \Gamma_i \Gamma_j) = \delta_{ij}$.
Without loss of generality, the basis can be selected as
$\left\{ \Gamma \right\} = \left\{
    I_{2}, \sigma_{x}, \sigma_{y}, \sigma_{z}
    \right\}^{\otimes 2}$,
where $\sigma_{x,y,z}$ are the Pauli matrices.

Therefore, $\rho$ can be expressed as a linear combination of $\Theta_{i}$ with $\Theta = (\theta_{1}, \theta_{2}, \cdots, \theta_{16})$.
\begin{equation}
    \rho = \sum_{i=1}^{16} \theta_{i} \Gamma_{i}
\end{equation}
For a given set of CG-POVM bases $\{ \Omega^{k}, I_{4} - \Omega^{k} \}_{k=1}^{n}$,
each CG-POVM operator $\Omega^{k}$ can also be expressed as a 16-dimensional vector
under the same bases $\{ \Gamma_{i} \}_{i=1}^{16}$:
\begin{equation}
    \Omega^{k} = \sum_{i=1}^{16} \omega^{k}_{i} \Gamma_{i}
\end{equation}
where $\omega^{k}_{i} = \Tr ( \Omega^{k} \Gamma_{i})$.
When measurements are performed using the CG-POVM bases,
the expectation values are given as:
\begin{equation}
    \hat{p}_{k} = \Tr( \rho \Omega^{k}) = \sum_{i=1}^{16} \theta_{i} \omega^{k}_{i}
    \label{eq:theory_result}
\end{equation}
leading to the linear equation,
\begin{equation}
    \begin{pmatrix}
        \omega^{1}_{1} & \omega^{1}_{2} & \cdots & \omega^{1}_{16} \\
        \omega^{2}_{1} & \omega^{2}_{2} & \cdots & \omega^{2}_{16} \\
        \vdots         & \vdots         & \ddots & \vdots          \\
        \omega^{n}_{1} & \omega^{n}_{2} & \cdots & \omega^{n}_{16} \\
    \end{pmatrix}
    \begin{pmatrix}
        \theta_{1}  \\
        \theta_{2}  \\
        \vdots      \\
        \theta_{16} \\
    \end{pmatrix}
    =
    \begin{pmatrix}
        \hat{p}_{0} \\
        \hat{p}_{1} \\
        \vdots      \\
        \hat{p}_{n}
    \end{pmatrix}
    \label{eq:theory_linear_equation}
\end{equation}
or simply, $X\Theta = \hat{P}$.
To solve above linear equation for two-qubit system, a minimum of 16 CG-POVM bases should be selected
to determine all elements of $\Theta_{i}$, that is $n=16$. The chosen set of CG-POVM bases $\{ \Omega^{k}, I_{4} - \Omega^{k} \}$ must ensure $\rank(X) = 16$,
corresponding to the full rank of the two-qubit $X$ matrix, enabling the solution of the linear equation.
By utilizing the Maximum Likelihood Estimation (MLE) technique to meet the physical constraints for the density matrix,
including hermiticity, positivity, and normalization, we can determine the optimal solution for $\Theta$
that satisfies these conditions\cite{qi2013quantum, smolin2012efficient}.

After constructing the CG-POVM basis set satisfying the full-rank condition,
we demonstrate the optimization of the CG-POVM basis set to minimize the impact of noise during measurement,
such as the shot noise which follows a Poisson distribution.
Let $\epsilon_{k} = \hat{p}_k - p_k$ represent the statistical error arising from the estimator $\hat{p}_{k}$,
which is proportional to $\sqrt{N_{p}}$, and $N_{p}$ is the total number of detected photons or electrons. From Equation (\ref{eq:theory_result}), we derive the following relation:
\begin{equation}
    \hat{p}_{k} = \sum_{i=1}^{16} \Theta_{i} \omega^{k}_{i} + \epsilon_{k}
\end{equation}
From Equation (\ref{eq:theory_linear_equation}), the reconstructed density matrix now incorporates the statistical error:
\begin{equation}
    \hat{\Theta} = X^{-1} \hat{P} = \Theta + X^{-1} \epsilon
\end{equation}
The magnitude of this error is contingent on $X$ and,
thus, on the selection of the set of CG-POVM bases $\{ \Omega^{k} \}_{k=1}^{16}$.
Let $X = U D U^{\dagger}$ represent the eigen-decomposition of $X$ where $D = \diag\{e_{1}, e_{2}, \cdots, e_{16}\}$ and $U$ is a unitary matrix.
The magnitude of the error is then:
\begin{equation}
    \left| \Theta_{\text{err}} \right|^{2} = \epsilon^{\dagger} (X^{\dagger} X)^{-1} \epsilon = \epsilon^{\dagger} U D^{-2} U^{\dagger} \epsilon.
    \label{eq:error_magnitude}
\end{equation}
where $D^{-1} = \diag\{e_{1}^{-1}, e_{2}^{-1}, \cdots, e_{16}^{-1}\}$.
We assume homogeneous error $\epsilon_{k}$ accross all measurement to pursue symmetric and information complete measurements. 
This harmonic mean of the eigenvalues of $X$ is minimized when the square of each eigenvalue of $X$ is uniform.
\begin{figure*}[t]
    \includegraphics[width=\textwidth]{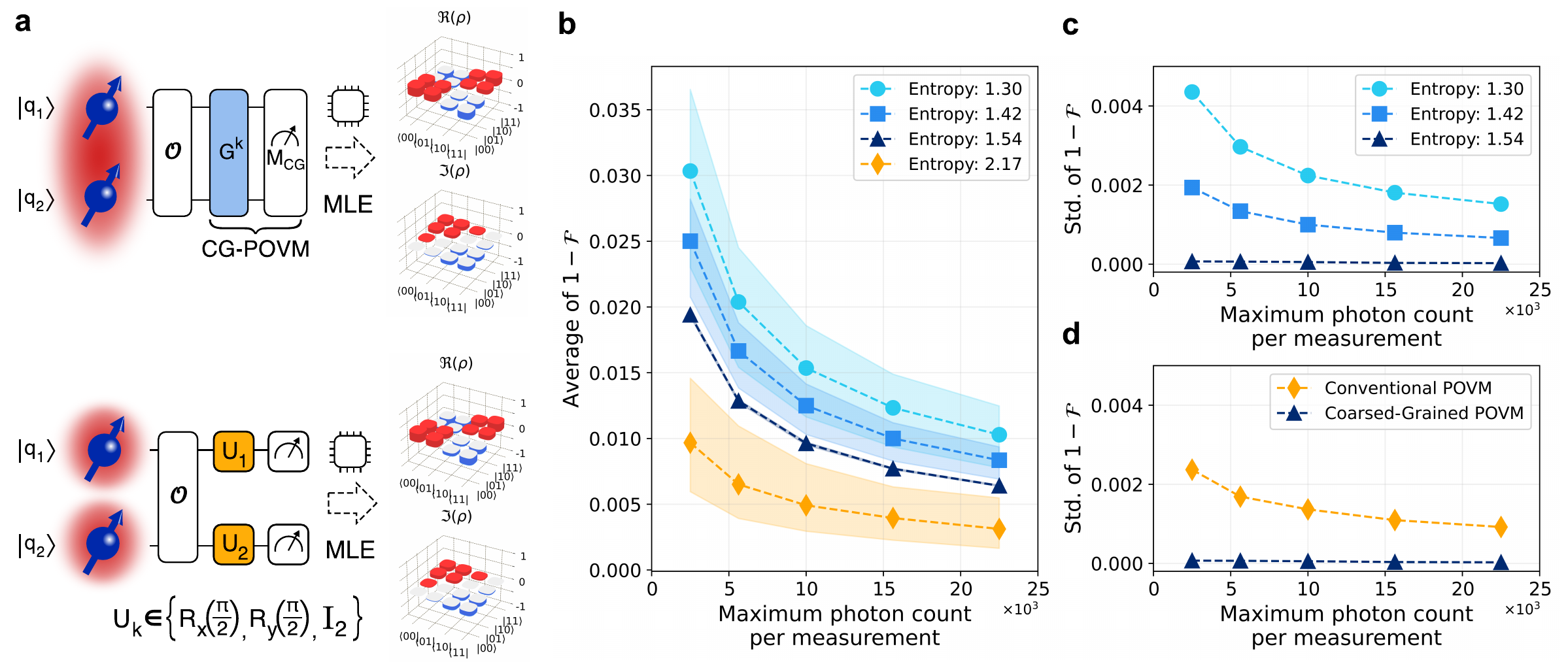}
    \caption{
        (a) Top: The CG-POVM QST process for a two-qubit system is depicted.
        Bottom: An individual readout scheme using conventional POVM basis.
        CG-QST and QST process is performed after the gate operation $\mathcal{O}$.
        (b) The graph presents the average infidelity $ 1 - F $ in relation to the maximum photon count for each measurement,
        highlighting a discernible trend where increasing von Neumann entropy of the POVM set correlates with a decrease in infidelity.
        This indicates that higher entropy POVM sets tend to be more effective in QST,
        demonstrating a more robust performance against noise and improved state reconstruction accuracy.
        (c) Standard deviation of infidelity $ 1 - F $ across target density matrices for various CG-POVM sets.
        The trend suggests that POVM sets with higher entropy tend to yield more consistent reconstruction results,
        reflecting their unbiased nature across a range of quantum states.
        (d) Comparison of the standard deviation of infidelity between the conventional POVM and the optimized CG-POVM sets.
        Despite the higher entropy of the conventional set, the optimized CG-POVM set closely approaches the performance of the GSIC POVM bases,
        potentially offering better resilience to noise and an unbiased estimation of quantum states.
    }
    \label{fig:fig2}
    
\end{figure*}
Our optimization approach involves initially deriving the Gram matrix from the POVM basis set,
followed by determining the eigenvalues, and subsequently computing the von Neumann entropy based on the extracted eigenvalues.
This strategy aims to mitigate the impact of statistical error on density matrix reconstruction
by minimizing $\Tr \left(D^{-1}\right)^{2}$ (in Eq.
(\ref{eq:error_magnitude})), as quantified through the Gram matrix representation.
The Gram matrix $\Pi$ is defined as $\Pi = X^{\dagger}X$, where each element is the inner product of the POVM bases:
\begin{equation}
    \Pi_{ij} = \Tr \Omega^{i} \Omega^{j}
    \label{eq:gram_matrix}
\end{equation}
The eigenvalues of the Gram matrix represent the influence of each measurement on a certain basis\cite{shawe2005eigenspectrum, renes2004symmetric}.
Next, the von Neumann entropy of the Gram matrix is introduced as a measure of its eigenvalue uniformity:
\begin{equation}
    S = - \Tr (\Pi^{\prime} \ln \Pi^{\prime}) = - \sum_{i=1}^{16} e_{i}^{\prime} \ln e_{i}^{\prime}
    \label{eq:von_neumann_entropy}
\end{equation}
where $\Pi^{\prime}$ is a normalized Gram matrix, $\Pi^{\prime} = \Pi / \Tr \Pi$ so that $\sum_{i=1}^{16} e_{i}^{\prime} = 1$. 
By computing the von Neumann entropy of the Gram matrix, we gauge the impartiality of the constructed CG-POVM basis set.
Opting for a CG-POVM basis set with greater von Neumann entropy facilitates a more uniform measurement over the entire Hilbert space,
contributing to a decrease in the statistical error of the estimated density matrix.
When the entropy reaches its maximum, the vector $\omega_{i}^{k}$ is uniformly spread throughout the Hilbert space,
ensuring that the basis elements of the CG-POVM are uniformly correlated.
Consequently, the CG-POVM basis set becomes close to the status of a GSIC POVM set (Supplementary Information). Please note that our optimal CG-POVM construction argument can be generalized to an N-qubit system with dimension $4^{N}$.

\subsection{Optimal CG-QST Simulation Results}

Figure \ref{fig:fig2}a illustrates the schematic diagram of the CG-QST process (upper) and the conventional individual readout QST process (lower) for two-qubit system. Applying proposed method to determine the optimal CG-POVM basis set,
we compare the infidelity of QST using our approach with the conventional individual readout scheme.
We conduct 100 Monte Carlo (MC) simulations of the QST process using the CG-POVM basis set
for each 36 symmetrically chosen target density matrices of a two-qubit system where the target states are eigenstates of Pauli matrices $\left\{\ketbra{\psi_{i}} \otimes \ketbra{\psi_{j}}\right\}$ and $\ket{\psi_{i}} \in \left\{ \ket{0}, \ket{1}, \ket{+}, \ket{-}, \ket{i}, \ket{-i} \right\}$,
incorporating randomly generated Poisson noise. The same number of MC simulations were conducted using the conventional Pauli matrix-based POVM basis set with selective readout. We used Pauli measurement basis $\{ \ketbra{\psi_{i}} \otimes \ketbra{\psi_{j}} \}$,
where $\ket{\psi_{i}}$ is one of the following states $ \left\{ \ket{0}, \ket{1}, \ket{+}, \ket{i} \right\}$.
The infidelity $1- F$ in QST is expressed as $1 - \Tr \sqrt{\sqrt{\rho} \hat{\rho} \sqrt{\rho}}$,
with $\rho$ representing the target density matrix and $\hat{\rho}$ denoting the reconstructed density matrix.
The mean infidelity value represents the each method's robustness against noise,
while the standard deviation of the infidelity gauges the method's independence from the target state.

As shown in Figure \ref{fig:fig2}b, in general,
the POVM basis set with higher von Neumann entropy exhibits superior performance in terms of mean infidelity, indicating greater robustness against statistical errors. In light of this, the conventional measurement operator set, having a von Neumann entropy of 2.17, still outperforms the CG-POVM basis set,
which achieves a maximum entropy of 1.54. Meanwhile, the difference in infidelity is sufficiently close,
suggesting that a comparable order of infidelity magnitude could be attained by just increasing the number of measurement shots.
Furthermore, the trend within CG-POVM basis sets with different entropy values suggests that higher entropy correspond to lower standard deviations of infidelities.
This indicates that a more uniform CG-POVM basis set leads to more unbiased reconstructions across different target density matrices (Fig.\ref{fig:fig2}c).
One important remark is that the optimal CG-POVM basis set, achieved by maximizing the von Neumann entropy to 1.54,
coincides the GSIC basis under the coarse-grained measurement condition (Supplementary Information).
This highlights the effectiveness of our scheme in constructing a nearly unbiased operator set.
For instance, although the conventional Pauli matrix-based POVM basis set, with higher entropy,
exhibits greater resilience to Poisson noise compared to the CG-POVM basis set, it lacks optimization for proximity to its GSIC.
Consequently, the unbiased nature of the optimized CG-POVM basis set might outperform the conventional POVM basis set (Fig.\ref{fig:fig2}d).
\begin{figure}[t]
\centering
    \includegraphics[width=0.44\textwidth, height=0.8\columnwidth]{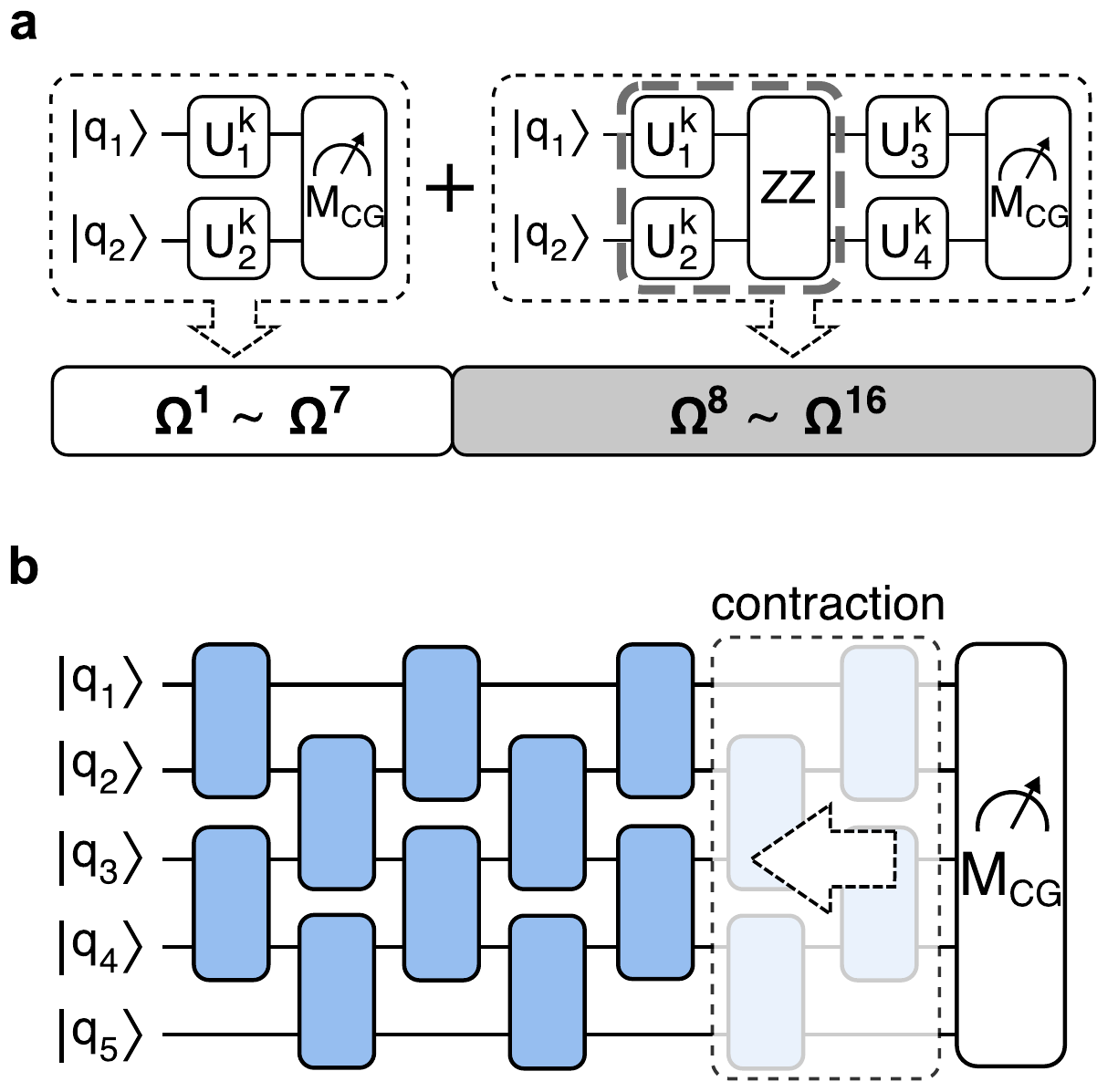}
    \caption{
        (a) Gate-efficient construction strategy for CG-POVM basis in a two-qubit system.
        The figure illustrates a two-tiered approach where gate blocks $ G^{k} $ are layered:
        the first seven $ \Omega^{1} $ to $ \Omega^{7} $ basis consist only of single-qubit gates (class-\uppercase\expandafter{\romannumeral1}),
        while the subsequent nine $ \Omega^{8} $ to $ \Omega^{16} $ involve a paramatrized Ising coupling gate $e^{-i\frac{\theta}{2} \sigma_{z} \otimes \sigma_{z}}$ (ZZ) encapsulated by single-qubit gates (class-\uppercase\expandafter{\romannumeral2}).
        (b) general multi-qubit quantum circuit with a brick-wall gate block structure for CG-POVM. 
        The circuit depth contraction scheme enables full Pauli basis measurement capability with complexity scaling that is non-exponential.
    }
    \label{fig:fig3}
\end{figure}
\subsection{Resource-Efficient CG-POVM Construction}
We propose a resource-efficient strategy for constructing a set of CG-POVM bases, aiming to minimize the utilization of two-qubit gates and the circuit depth.

First, we compose each gate block $G^{k}$ with two different classes:
a class-\uppercase\expandafter{\romannumeral1} consists solely of single-qubit gates, and class-\uppercase\expandafter{\romannumeral2} contains one Ising coupling gate encapsulated by single-qubit gates (Fig.\ref{fig:fig3}a).
For the example of two qubit system, the resultant unitary gate $G^{k}$ is then in $\mathbb{SU}(2) \otimes \mathbb{SU}(2)$ for class-\uppercase\expandafter{\romannumeral1} and $\mathbb{SU}(4)$ for class-\uppercase\expandafter{\romannumeral2}.
Then, we initiate the construction of a set of CG-POVM bases by layering $G^{k}$ exclusively with class-\uppercase\expandafter{\romannumeral1},
sequentially transitioning each $G^{k}$ block from class-\uppercase\expandafter{\romannumeral1} to \uppercase\expandafter{\romannumeral2} until meeting the full-rank condition.
Although the GSIC condition is not satisfied now, this strategy guarantees the full-rank condition of the $X$ matrix with suitable parameters for the quantum circuit ansatz,
while minimizing the number of required two-qubit gates.
Minimizing the number of two-qubit gates is crucial, as they tend to introduce more noise compared to single-qubit gates,
affecting the overall fidelity of generating CG-POVM bases.
For the example of two-qubit system, our strategy yields 7 CG-POVM bases, $1 \leq k \leq 7$,
without utilizing any two-qubit gate and the remaining 9 CG-POVM bases, $8 \leq k \leq 16$,
with the inclusion of one two-qubit gate, thus constructing fully ranked CG-POVM bases (Fig.\ref{fig:fig3}a). Minimum required 9 class-\uppercase\expandafter{\romannumeral2} $G^{k}$ gate blocks corresponds to the number of joint Pauli bases in the conventional selective readout case.
Following our strategy of minimizing the number of two-qubit gates,
we can further optimize CG-POVM basis set by maximing the von Neumann entropy of the Gram matrix, using only 9 class-\uppercase\expandafter{\romannumeral2} $G^{k}$ gate blocks out of 16 bases. Under these conditions, the highest entropy achieved is 1.516,
slightly lower than the maximum entropy achievable with CG-POVM GSIC bases constructed using all 16 bases with class-\uppercase\expandafter{\romannumeral2} $G^{k}$ gate blocks.
Nevertheless, comparable QST performance is expected for the gate efficient CG-POVM basis set.

Next, for the $N>2$ multi-qubit system, we contract the number of layers from the large circuit layer until the CG-POVM operator can no longer measure the entire set of Pauli bases.
In general, gate blocks in quantum circuit can be arranged in a brick-wall structure with all gate types of class-\uppercase\expandafter{\romannumeral2}, encompassing two-qubit gates (Fig.\ref{fig:fig3}b). At first, a sufficiently large number of CG-POVM operators are generated for a given circuit ansatz and the number of circuit layers,
employing random variables for the circuit parameters, to ensure that the vector $\omega_{i}^{k}$ is distributed throughout the entire Hilbert space. We then sequentially contract the circuit layer, verifying if the CG-POVM operator can measure a specific Pauli basis by computing $\omega_{i}^{k}$ and checking for nonzero values in $\omega_{i}^{k}$. If $\omega_{i}^{k} = 0$ for any $i$ across all $k$, it indicates an inability to extract information for that particular Pauli basis, resulting in the failure of QST. This confirmation approach is more feasible for contracting an N-qubit system compared to directly determining the rank of the CG-POVM set, which requires significantly more computational intensity. Thus, our approach enables us to determine the minimal number of layers required for the CG-POVM within this quantum circuit framework to measure all Pauli bases.

\subsection{Multi qubit CG-QST Simulation Results}
\begin{figure}[t]
    \includegraphics[width=0.48\textwidth]{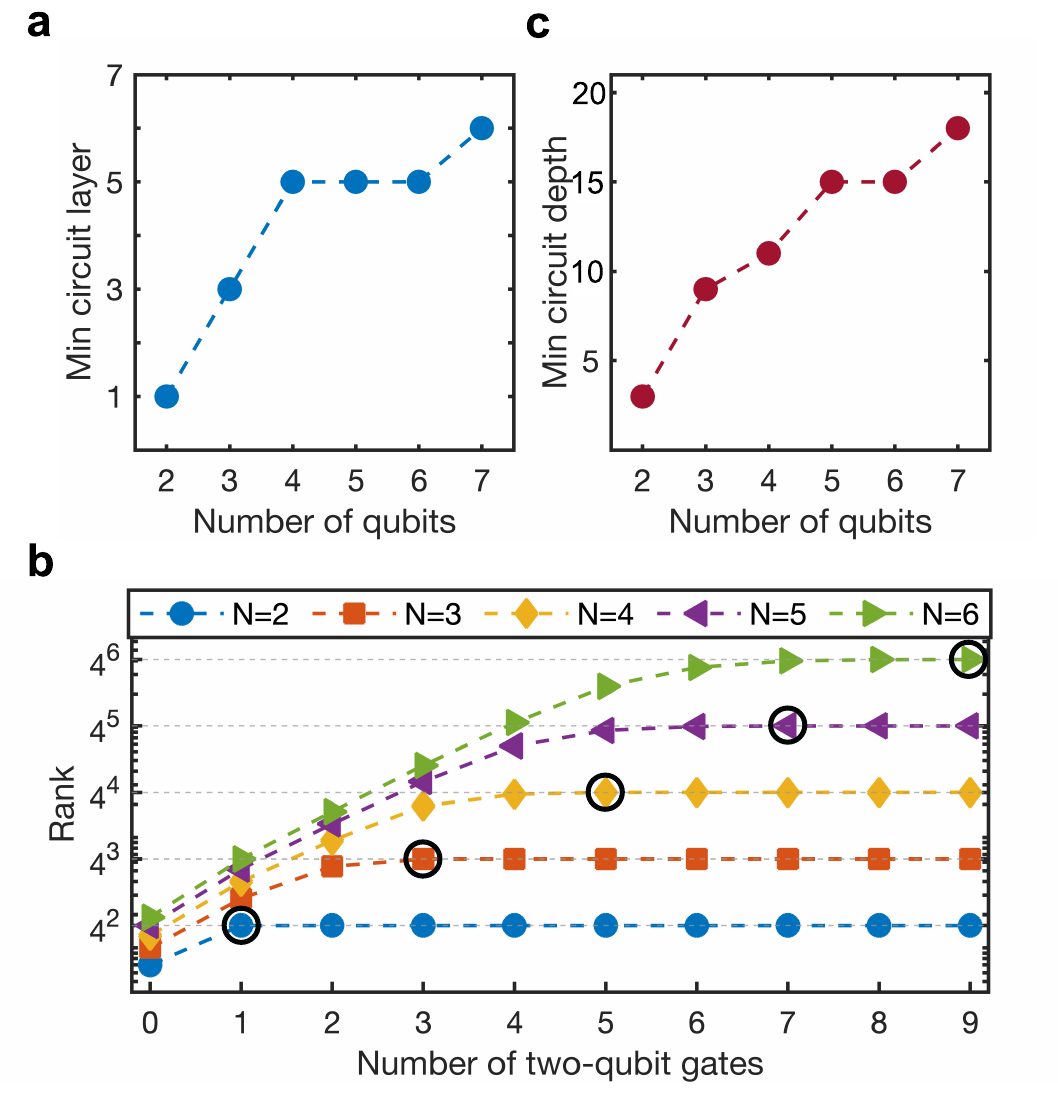}
    \caption{
        (a) The relation between the minimum circuit layers and the number of qubits, showing non-exponential scaling for full Pauli basis measurement.
        (b) The relationship between the number of two-qubit gates and the rank of the $ X $ matrix, indicating efficient scaling of CG-POVM respect to total qubit number N.
        (c) Minimum circuit depth as a function of qubit number, affirming the scalability of CG-POVM construction without exponential growth in depth.
    }
    \label{fig:fig4}
\end{figure}

Utilizing the aforementioned resource-efficient strategy, we demonstrate that our CG-POVM construction scheme for QST can be extended to an $N$-qubit system without exponential increases in the number of circuit layers, the number of two-qubit gates per measurement, and the circuit depth.
The number of circuit layers corresponds to the quantity of gate block columns in the quantum circuit ansatz,
while the circuit depth is defined as the maximum path length within the circuit.
Specifically, the $G^{k}$ class-\uppercase\expandafter{\romannumeral1} gate block has a circuit depth of 1, whereas class-\uppercase\expandafter{\romannumeral2} has a circuit depth of 3.

First, Figure \ref{fig:fig4}a illustrates the minimum number of layers required as the number of qubits increases, using the circuit layer contraction strategy introduced in the previous section.
The result indicates that the minimum number of layers does not increase exponentially with the number of qubits. Next, it is important to note that the specific Pauli bases can be constructed with the CG-POVM basis set
without necessitating the inclusion of all gate block types with class-\uppercase\expandafter{\romannumeral2}.
The rank calculation is conducted after determining the minimum circuit layer for various numbers of qubits. 
Figure \ref{fig:fig4}b depicts the maximum achievable rank of the $X$ matrix by varying the number of two-qubit gates per CG-POVM operator. The full-rank condition of the $X$ matrix is satisfied when the number of two-qubit gates per measurement is $2N-3$.
Thus, we confirm that the number of two-qubit gates per measurement also does not grow exponentially with the number of qubits. Finally, as shown in Figure \ref{fig:fig4}c,
the minimum required circuit depth is computed by varying the number of qubits.
For a fixed number of qubits, we utilize a minimized number of circuit layers
with the minimum number of $2N-3$ two-qubit gates per CG-POVM operator,
constructing all possible quantum circuits for the minimum depth calculation.
The circuit depth minimization process follows the same strategy, validating whether the CG-POVM operator can measure the entire Pauli basis, as used in circuit layer minimization. We also confirm that the minimum circuit depth required does not increase exponentially with the number of qubits.
After constructing CG-POVM basis set with the most resource-efficient circuit,
we can further optimize the set to minimize the effect of noise during measurement
by increasing the von Neumann entropy of the Gram matrix as illustrated in Figure \ref{fig:fig2}.

\subsection{Conclusion}
In summary, the CG-POVM basis set, characterized by a $4^N$-rank $X$ matrix, supports QST for an $N$-qubit system, yielding $4^N$ measurement outcomes.
The MLE technique is employed to determine the optimal solution for the target density matrix $\rho$ while adhering to physical constraints.
Additionally, optimization of the CG-POVM basis set, aimed at minimizing the impact of noise during measurement,
involves maximizing the von Neumann entropy of the Gram matrix. With an increasing number of qubits, the growth of two-qubit gates per measurement
and circuit depth avoids exponential escalation, making CG-QST an efficient approach for N-qubit systems.
We anticipate the applicability of our strategy not only in constructing CG-QST but also in designing target set of POVM bases with CG constraints,
leveraging the universality of parameterized quantum circuit ansatz methodology\cite{benedetti2019parameterized, haug2022natural, PhysRevResearch.2.033125}.

Designing an optimal CG-QST scheme can be directly applied to solid-state embedded defect qubit systems
such as the nitrogen-vacancy (NV) centers or the silicon-vacancy (SiV) centers in diamond\cite{PhysRevB.79.235210, sukachev2017silicon}.
One of the challenges in achieving a scalable electronic spin defect-based qubit system is the lack of individual readout capability.
The proposed CG-QST technique offers a promising solution enabled by the individual control of each spin defect
through different quantization axes\cite{dolde2013room, lee2023dressed} or the same quantization axes with frequency encoding techniques\cite{zhang2017selective} and engineered two-qubit gates via dipolar interactions\cite{lee2023dressed}.
Extending beyond solid-state qubits, the CG-QST method presents an opportunity to create a more resource-efficient multi-qubit platform
by reducing the number of detectors required compared to the total number of qubits\cite{lib2023resourceefficient}.
In our specific implementation, the CG-POVM basis set necessitated only one detector for an $N$-qubit state reconstruction,
distinguishing only two states, such as signal-in-state and no-signal-state.
In the general scenario, the number of detectors can exceed one, and the count of distinguishable states can surpass two as well.
This leads to a higher von Neumann entropy of the Gram matrix, thereby enhancing the performance of QST.
Hence, a trade-off emerges between the quantity of detectors and QST performance.
Nevertheless, a comparable range of fidelity can be always attained through an increase in the measurement shot number.
The resource-efficient optimal CG-POVM construction presents an alternative avenue, offering minimized detector usage
while maximizing measurement performance for multi-qubit platforms.

\subsection{Acknowledgement}
The authors would like to thank Yong Siah Teo, and Conlon Lorcan for helpful discussions. This material is based upon work supported by, or in part by, the KIST institutional program (Project. No. 2E32941) and the National Research Foundation of Korea STEAM program under Grant No. 2N73600, and A*STAR (Project No. C230917009).
\subsection{Contributions}
J.L., Y.C. and Y.K. conceived the idea, J.L. and Y.K. co-supervised the project. D.J. developed and performed numerical simulations. All authors analyzed and discussed the data. All authors participated in writing the manuscript. 

\subsection{References}
\bibliography{reference.bib}

\end{document}